\documentclass[12pt,twoside]{article}   
\usepackage[super,sort,comma]{natbib}

\usepackage{fancyhdr}		

\usepackage{amsmath}
\usepackage{upgreek}

\newcommand{\captionv}[3]{\begin{center}\parbox{#1cm}{\caption[#2]{{\sf #3}}}
        \end{center}}

\usepackage[section]{placeins}   %

\usepackage{graphicx}

\makeatletter \renewcommand\@biblabel[1]{$^{#1}$} \makeatother
\newcommand{\cen}[1]{\begin{center} #1 \end{center}}

\usepackage[mathlines]{lineno}

\usepackage{hyperref}
\hypersetup{ colorlinks,
    citecolor=blue,
    filecolor=blue,
    linkcolor=blue,
    urlcolor=blue
}

\usepackage{xcolor}

\definecolor{gray}{rgb}{0.6,0.6,0.6}
\definecolor{red}{rgb}{0.85,0,0}
\definecolor{green}{rgb}{0,0.85,0}
\definecolor{blue}{rgb}{0,0,0.85}
\definecolor{beige}{rgb}{0.92,0.87,0.78}
\usepackage[all]{hypcap}    

\usepackage{multirow}

\usepackage{makecell}

\begin{document}

\cen{\sf {\Large {\bfseries Evaluation of the two-voltage method for parallel-plate ionization chambers irradiated with pulsed beams} \\  
\vspace*{10mm}
José~Paz-Martín\textsuperscript{1*},
Andreas~Schüller\textsuperscript{2},
Alexandra~Bourgouin\textsuperscript{2**},
Araceli Gago-Arias\textsuperscript{1},
Diego~M.~González-Castaño\textsuperscript{3},
Nicolás~Gómez-Fernández\textsuperscript{3},
Juan~Pardo-Montero\textsuperscript{4,5},
Faustino~Gómez\textsuperscript{1,3}
}\\
\textsuperscript{1} Departamento de Física de Partículas, Universidade de Santiago de Compostela, Santiago de Compostela, 15782, A Coruña (Spain).\\
\textsuperscript{2} Physikalisch-Technische Bundesanstalt, Bundesallee
100, Braunschweig, 38116, (Germany).\\
\textsuperscript{3} Laboratorio de Radiofísica, Universidade de Santiago, Estrada de San Lourenzo, Santiago de Compostela, 15782, A Coruña, (Spain).\\
\textsuperscript{4} Group of Medical Physics and Biomathematics, Instituto de Investigacion Sanitaria de Santiago (IDIS), Travesía da Choupana, Santiago de Compostela, A Coruña, 15706, (Spain).\\
\textsuperscript{5} Department of Medical Physics, Complexo Hospitalario Universitario de Santiago de Compostela, Travesía da Choupana, Santiago de Compostela, A Coruña, 15706, (Spain).\\
\vspace{5mm}
Version typeset \today\\
}
\pagenumbering{roman}
\setcounter{page}{1}
\pagestyle{plain}

\textsuperscript{*} Author to whom correspondence should be addressed. Email: jose.martin@usc.es \\

\textsuperscript{**} Current affiliation: National Research Council of Canada, Metrology Research Center, 1200, Montreal Road, Ottawa, K1A0R6, ON, Canada. \\

\begin{abstract}
\noindent {\bf Background:} Air-vented ionization chambers exposed to clinical radiation beams may suffer from recombination during the drift of the charge carriers towards the electrodes. Thus, dosimetry protocols recommend the use of a correction factor, usually denominated saturation factor ($k_{\rm sat}$), to correct the ionization chamber readout for the incomplete collection of charge. The two-voltage method is the recommended methodology for the calculation of the saturation factor, however, it is based on the early Boag model, which only takes into account the presence of positive and negative ions in the ionization chamber and does not account for the electric field screening or the free electron contribution to the signal.\\

\noindent {\bf Purpose:} To evaluate the impact of a more realistic approach to the saturation problem that accounts for the free electron fraction.\\

\noindent {\bf Methods:} The saturation factor of four ionization chambers (two Advanced Markus and two PPC05) was experimentally determined in the ultra-high dose per pulse reference beam of the German National Metrology Institute (Physikalisch-Technische Bundesanstalt, PTB) for voltages ranging from 50~V to 400~V and pulse durations between 0.5~$\upmu$s and 2.9~$\upmu$s. Several analytical models and a recently developed numerical model are used to calculate the saturation factor as a function of the dose per pulse and compare it to the obtained experimental data. Parametrizations of the saturation factor against the ratio of charges at different voltages are given for PPIC with a distance between electrodes of 0.6~mm and 1~mm in pulsed beams for different pulse durations. \\
 
\noindent {\bf Results:} The saturation factors calculated using the different Boag analytical models do not agree neither with each other nor with the numerical simulation even at the lowest dose per pulse of the investigated range ($<$~30~mGy). A recently developed analytical model by Fenwick and Kumar agrees with the numerical simulation in the low dose per pulse regime but discrepancies are observed when the dose becomes larger (i.e. $>$~40 mGy for Advanced Markus) due to the electric field perturbation. The numerical simulation is in a good agreement with the experimentally determined charge collection efficiency with an average discrepancy of 0.7~\% for the two PPC05 and 0.5~\% for the two Advanced Markus. The saturation factor obtained with the numerical simulation of the collected charge has been fitted to a third-order polynomial for different voltage ratios and pulse duration. This methodology provides a practical way for $k_{\rm sat}$ evaluation whenever $k_{\rm sat}<1.05$.\\ 

\noindent {\bf Conclusions:} The numerical simulation shows a better agreement with the experimental data than the current analytical theories in terms of charge collection efficiency. The classical two-voltage method, systematically overestimates the saturation factor, with differences increasing with dose per pulse but also present at low dose per pulse. These results may have implications for the dosimetry with ionization chambers in therapy modalities that use a dose per pulse higher than conventional radiotherapy such as intraoperative radiotherapy but also in conventional dose per pulse for ionization chambers that suffer from significant charge recombination.

\end{abstract}

\newpage     

\tableofcontents

\newpage

\setlength{\baselineskip}{0.7cm}      

\pagenumbering{arabic}
\setcounter{page}{1}
\pagestyle{fancy}

\section{Introduction}
Air-vented ionization chambers are the gold standard for determining the absorbed dose to water in external beam radiotherapy. Under reference conditions\citep{agency_absorbed_2024}, the readout of an ionization chamber may be converted to absorbed dose to water by using the calibration coefficient alongside several correction factors addressing the impact of temperature, pressure and polarity. Furthermore, when the beam quality differs from that used for calibration, a beam quality correction factor is included.

Among the correction factors, the saturation factor typically denoted as $k_{\rm sat}$ or $P_{\rm ion}$, accounts for the incomplete collection of charge resulting from the recombination of charge carriers. Occasionally, the inverse of the saturation factor, commonly referred to as charge collection efficiency (CCE), can be reported instead. The two-voltage method (TVM) is the recommended methodology by the TRS-398 and TG-51 codes of practice\citep{agency_absorbed_2024, almond_aapms_1999} to determine the saturation factor for ionization chambers. In the TVM, based on Boag's formalism\cite{boag_ionization_1950}, for a given collected charge ratio and a certain voltage ratio, the saturation factor is uniquely determined. Therefore, there is no dependency of the collected charge ratios on the geometry or other parameters involved in the physical problem, such as ion mobilities or the ion-ion recombination coefficient. This method, based on the early Boag model\cite{boag_ionization_1950}, considers the ionization chamber electric field unperturbed and only two charge carriers inside the gas volume, namely: positive and negative ions.

Other analytical theories developed by Boag and colleagues include the free electrons that do not attach to neutral molecules in air, using an approximation of the electron density distribution along the ionization chamber\citep{boag_effect_1996}. More recently, Fenwick and Kumar\citep{fenwick_collection_2022} obtained an analytical formula that takes into account the exact distribution of electrons inside a parallel plate ionization chamber (PPIC), providing the most complete analytical picture up to this moment. However, the analytical models do not take into account effects such as the electric field perturbation or the pulse duration, which can be significant when the saturation factor becomes large. 

Alternatively, numerical approaches have been applied under high-dose and ultra-high dose per pulse beams showing satisfactory results\cite{gotz_new_2017, kranzer_ion_2021, gomez_development_2022, paz-martin_numerical_2022, kranzer_charge_2022}. This method allows the description of the recombination, electron attachment, and space-charge effects during the drift of the charge carriers using a set of partial differential equations discretized in space and time for each charge carrier considered. On the other hand, empirical descriptions such as the Petersson \textit{et al.} \citep{petersson_high_2017} and Bourgouin \textit{et al.}\citep{bourgouin_charge_2023} logistic fits for different PPICs models have been proposed as well as a semi-analytical approach from Bancheri and Seuntjens\citep{Bancheri_A_2024}.

In this work, the CCE of two commercial PPIC models has been studied for different voltages and DPPs up to 460~mGy per pulse. After a brief comparison between the analytical models against the numerical model in Section~\ref{S_comp}, the experimentally determined CCE in an ultra-high dose per pulse electron beam is evaluated in Section~\ref{S_CCE} and compared against the predictions of a recently developed numerical model. Also, the polarity effect is evaluated and the CCE is compared with a previous study carried out by Bourgouin \textit{et al.}\citep{bourgouin_charge_2023}. In Section~\ref{S_TVM} the experimental charge ratios are compared to the classical TVM and the numerical model. Finally, in Section~\ref{S_fit} the coefficients from a polynomial fit of the saturation factor against the collected charge obtained from the simulation are given for several voltage ratios and beam pulse durations (0.5~$\upmu$s~-~5.0~$\upmu$s).

\section{Methods and materials}

\subsection{Investigated ionization chambers}
For this investigation, two commercial PPIC models were considered, namely: Advanced Markus and  PPC05, with a nominal distance between electrodes of 1.0~mm and 0.6~mm, respectively. For each ionization chamber model, two samples were investigated. Both Advanced Markus and one PPC05 were calibrated at the reference $^{60}$Co of the PTB with a relative uncertainty of 0.25~\% (k~=~1) while one PPC05 was calibrated in the $^{60}$Co source of the Radiation Physics Laboratory at the University of Santiago de Compostela with a relative uncertainty of 0.40~\% (k~=~1). For the two PPC05 ionization chambers, the distance between electrodes was also measured using a micro-CT.\\

During the experimental campaign, the high voltage was supplied to the high voltage electrode (upper electrode) for the Advanced Markus model while it was supplied to the collecting electrode and guard ring for the PPC05. Thus, when the applied voltage is positive, the collected charge polarity is positive in the Advanced Markus PPIC and negative in the PPC05. The parameters related to the PPIC used in this paper for the analysis of the data and the numerical calculations are listed in Table~\ref{ppic_parameters}.
\begin{table}[!t]
\begin{center}
\vspace*{2ex}
\captionv{15}{}{List of the principal parameters of the investigated PPICs used in this work. The nominal sensitive radius, recommended voltage, and maximum voltage are obtained from the corresponding manufacturer's datasheet. All the ionization chambers were calibrated at 300~V.\label{ppic_parameters}\vspace*{2ex}}
\begin{tabular} {lcccccc}
\hline
\hline
 & & \multicolumn{2}{c}{\textbf{PPC05}}  & & \multicolumn{2}{c}{\textbf{Advanced Markus}} \\
\cline{3-4} \cline{6-7}
 & S/N & 1178 & 1496 & & 2320 & 2350 \\
\hline
Calibration coefficient (Gy\,nC$^{-1}$)  & & 0.5812(14) & 0.5881(24) & & 1.4413(36) & 1.4867(37)     \\
Measured gap (mm)  & & 0.648 & 0.612 &  & - & - \\
$k_{\text{Q}}$  & &  \multicolumn{2}{c}{0.8918(62)}  &  &  \multicolumn{2}{c}{0.9014(63)} \\
Nominal gap (mm)  & &   \multicolumn{2}{c}{0.6}  &  & \multicolumn{2}{c}{1} \\
Nominal sensitive radius (mm) & & \multicolumn{2}{c}{5}  &  & \multicolumn{2}{c}{2.5} \\
Recommended voltage (V) & & \multicolumn{2}{c}{$\pm$ 300}  &  & \multicolumn{2}{c}{$\pm$ 300} \\
Maximum voltage (V) & & \multicolumn{2}{c}{$\pm$ 500}  &  & \multicolumn{2}{c}{$\pm$ 400} \\
\hline
\hline
\end{tabular}
\end{center}
\end{table}
\subsection{Experimental setup}
The experimental campaign was performed in the ultra-high pulse dose rate reference electron beam\citep{bourgouin_characterization_2022} of the metrological electron accelerator facility\citep{schuller_metrological_2019} (MELAF) of the Physikalisch-Technische Bundesanstalt (PTB) at Braunschweig, Germany.

The PPICs were irradiated using an electron beam with an energy of 20~MeV. To achieve the dose per pulse (DPP) range of interest of this investigation, 3 plates of a total thickness of 6~mm made of aluminum were placed after at the beam exit window of the accelerator in order to scatter the electrons and to form a large field. The chambers were placed at the reference point of measurement in the water phantom, which has an entrance window made of 5.62~mm PEEK. The water phantom was placed at a source-to-surface distance of 90 cm measured from beam exit window of the beam line. In this investigation, the pulse  duration of the beam was varied between 0.5~$\upmu$s up to 2.9~$\upmu$s and the DPP was changed by means of a slit opening in the beam line of the linear accelerator\citep{bourgouin_characterization_2022}, with an aperture between 2~mm and 8~mm. The linear accelerator is equipped with an in-flange Integrating Current Transformer (ICT) (Bergoz) which measures the charge per pulse\cite{schuller_metrological_2019} of the beam and is used to correct for pulse-to-pulse variations. Typically, the ICT measured charge per pulse ranges from 20~nC to 300~nC and a statistical relative uncertainty of 0.1~\% with an absolute uncertainty of 0.015~nC were considered\citep{bourgouin_charge_2023}. The pulse duration was measured as the full width at half maximum. Throughout all the measurements the beam pulse repetition frequency was 5~Hz.

The PPICs were connected to a Keithley 616 electrometer (Keithley Instruments) and measurements were performed in current mode. Additionally, a 33~nF capacitor was also connected in parallel to the electrometer input to avoid any possible deviation from linearity due to the high voltage at the electrometer input. A calibration factor for the electrometer $k_{elec}$ was determined using a reference standard constant current source and applied to correct the electrometer readout of the collected charge. The PPICs were polarized using an in-house PTB high-voltage source from 50~V to 400~V in steps of 50~V for both, positive and negative polarity. The high voltage output was measured using a digital voltmeter. All the measurements were acquired sequentially, recording 25 beam pulses for each value of voltage, DPP and pulse duration. The standard deviation of the PPIC collected charge corrected using the ICT readout was employed as a Type-A uncertainty estimation.

The half-value depth $R_{50}$ of the electron beam was determined using the depth dose curve measured with a flashDiamond (PTW T60025)\citep{marinelli_design_2022, kranzer_response_2022} (in the following, fD) that its response is independent of both DPP and intra pulse dose rate in the range of interest for this work. The reference depth of measurement for positioning the ionization chambers was determined following the TRS-398 code of practice\citep{agency_absorbed_2024}. The PPICs were positioned at the reference depth of measurement accounting for the water-equivalent thickness of the entrance window.

\subsection{Determination of the absorbed dose to water}

The reference dosimetry was performed using alanine pellets in combination with the fD. The alanine measurements were used to determine a calibration coefficient of the fD and then, the fD was used to obtain the DPP in each of the configurations used.

The alanine dose measurements were converted to DPP delivered by the electron beam using the scaling factor determined by Vörös \textit{et al}\citep{mvoros_relative_2012} of 1.012(10). Additionally, a field correction factor that accounts for the beam radial non-uniformity was applied\citep{bourgouin_absorbed-dose--water_2022}. In total, 9 points with different slit opening and pulse duration were used to determine the calibration coefficient of the fD.

For each slit opening, the dose per pulse was varied by changing the pulse duration. A 3rd order polynomial fit of the DPP versus the charge per pulse measured by the ICT was performed. This polynomial fit addresses the non-linear relationship observed of the ICT signal with the absorbed dose while changing the pulse duration of the radiation delivery\citep{bourgouin_the_2023}. The ICT was also used to correct for any variation in the actual beam charge delivered per pulse.

All the measurements presented were performed approximately in one week. The charge per pulse measurements for the polynomial fit were done on the first day and the alanine calibration was performed at the end of the experimental campaign in which the polynomial fit's stability was also verified. Moreover, the standard deviation of the differences between the charge per pulse predicted by the polynomial fits and the measured charge per pulse from the fD at the end of the campaign was used as an estimation of the short-term stability uncertainty.

\subsection{Determination of the charge collection efficiency and the polarity factor}
The CCE was determined as the quotient of the DPP reading obtained with the PPICs and the actual DPP determined by the reference dosimetry, $D_{\text{ref}}$,

\begin{equation}
\text{CCE} = \frac{Q\,k_{\text{elec}}\,k_{\text{TP}} \,k_{\text{Q}, \text{Q}_0}\,N_{\text{D,w,Q}_0}}{D_{\text{ref}}}; \quad\quad\quad Q = \frac{|Q_{+}| + |Q_{-}|}{2},
\end{equation}

\noindent where $Q_{+}$ and $Q_{-}$ are the collected charge at positive or negative voltage, respectively, $k_{\text{elec}}$ is the electrometer calibration factor, $k_{\text{TP}}$ is the pressure and temperature correction of the ideal gas law, $N_{\text{D,w,Q}_0}$ the calibration coefficient of the ionization chamber, and $k_{\text{Q}, \text{Q}_0}$ is the beam quality correction.

\noindent The polarity correction factor was calculated according to the TRS-398 definition:
\begin{equation}
k_{\text{pol}} = \frac{|Q_{+}| + |Q_{-}|}{2|Q_{+}|}
\label{kpol_eq}
\end{equation}

\noindent The beam quality correction factors for the used PPICs were evaluated using the fits from the publication of Muir \textit{et al.}\citep{muir_monte_2014} using the $R_{50}$ measured with the fD. A relative uncertainty of 0.7 \% for the beam quality correction factor was considered, similar to Bourgouin \textit{et al.}\citep{bourgouin_charge_2023}.

\subsection{Numerical model\label{NumModel}}
A 1D numerical model, recently described in a publication of Paz-Martin \textit{et al.}\citep{paz-martin_numerical_2022}, was used to simulate the CCE of the investigated PPICs. This numerical model simulates in detail the attachment, multiplication, and recombination of the charge during its drift toward the electrodes. Also, the electric field perturbation due to the imbalance of charge is considered by solving the Poisson equation in each step. In the present study, the relevant parameters used for the transport coefficients are listed in Table~\ref{trasp_parameters}. The electron velocity and the electron attachment rate were simulated using the Magboltz\citep{biagi_monte_1999} simulation code (version 11.14). This code use the electron-atom (molecule) collision cross section to determine the electron swarm parameters for a given gas mixture using the monte-carlo method.

\begin{table}[!b]
\begin{center}
\vspace*{2ex}
\captionv{10}{}{Transport parameters used in the numerical simulations of this work.\label{trasp_parameters}\vspace*{2ex}}
\begin{tabular} {p{5 cm}cp{8 cm}}
\hline
\hline
\textbf{Transport parameter} & & \textbf{Description} \\
\hline
Electron velocity & & Function of the electric field, simulated using Magboltz. \\
Electron attachment rate & & Function of the electric field, simulated using Magboltz. \\
Ion mobilities & & From Zhang \textit{et al.}\citep{zhang_prediction_2019} \\
Ion-ion (volume) recombination coefficient & &  1.17$\times$10$^{-12}$~m$^3$\,s$^{-1}$\\
Electron-ion recombination coefficient & & Not used. \\
Electron multiplication coefficient & & Not used.  \\
Temporal structure & & Pulsed beam with a rectangular shape.\\
Spatial discretization & & 0.5~$\upmu$m. \\
Time discretization & & Adaptative, from 0.01~ns to 30~ns. \\
\hline
\hline
\end{tabular}
\end{center}
\end{table}

In line with previous publications \cite{gomez_development_2022}, electron-ion recombination's contribution is disregarded as it was estimated to be always below 0.2~\% and 0.1~\% on average for the DPP range studied and for all the chambers investigated. Although further investigation is needed, no significant differences in terms of CCE were found when enabling the electron multiplication process using tabulated first Townsend coefficient obtained from Magboltz\citep{biagi_monte_1999}. This is consistent with the fact that the nominal electric field of the chambers considered here is not high enough to trigger electron multiplication. The value used for the ion-ion recombination parameter is the value that reproduces better the experimental data presented in this work. 

It is worth mentioning that, although the electron attachment lifetime used in previous publications was taken from Boissonnat \textit{et al.}\cite{boissonnat_measurement_2016}, in this work we use the values obtained from Magboltz code, as they show slightly better agreement with the experimental data.

Some typical beam waveforms were recorded at PTB and introduced in the numerical model to evaluate their influence on the CCE. Despite local relative CCE deviations being up to 0.35~\%, on average the effect was always lower than 0.1~\%. Therefore, in this work, the fluence rate delivered to the chamber was considered a to be constant, i.e. rectangular temporal pulse shape.

When comparing with the experimental data, pulse duration, temperature, and pressure registered in each measurement point were introduced in the simulation. A relative humidity value of 50~\% was used in all simulations. During the analysis of the experimental data, differences between the absolute homologous voltages in positive and negative polarity up to a maximum of 7~V were noticed. The effect of this difference was estimated to be always below 0.1~\% in terms of CCE when the average of the positive and negative voltage is attributed to the polarity-corrected charge per pulse. Exhaustive numerical model
uncertainty evaluation has not been addressed in the present work due to the complexity of the analysis related to the different physical parameters and their correlations.

\subsection{Analytical models\label{AModels}}
Boag derived in his early work dealing with the description of volume recombination in ionization chambers the following expression to account for the charge collection efficiency CCE$_{\text{I}}$ for a PPIC  \citep{boag_ionization_1950}:

\begin{equation}
\text{CCE}_{\text{I}} = \frac{\log(1+u)}{u};\quad \quad u = \frac{\alpha \; N_0 \; d^2}{(\mu_+ + \mu_-) \; U }
\end{equation}

\noindent where  $N_0$ represents the charge carrier density of either sign released inside the chamber, $\alpha$ is the volume recombination coefficient, $d$ is the distance between electrodes, $U$ is the voltage applied and $\mu_{\pm}$ the positive and negative ion mobilities, respectively. Boag's formula is based on the assumptions of zero attachment time (electrons are instantaneously bounded into negative ions), no perturbation of the electric field across the chamber and all charges are released instantaneously (pulse duration is equal to zero).\\

Subsequently, three models were developed by Boag \textit{et al.}\citep{boag_effect_1996} to account for the free electrons when calculating the CCE. Considering a constant drift electron speed $v_e$ inside the chamber and a fixed electron attachment lifetime $\tau$, the Boag free electron fraction $p$ is defined as

\begin{equation}
p =\frac{Q_e}{Q_0} = \frac{v_e \tau}{d}\left(1-e^{-\frac{d}{v_e \tau}}\right);
\label{Boag_fef}
\end{equation}

\noindent where $Q_e$ is the total free electron charge arriving at the positive electrode and $Q_0$, is the total charge of either sign escaping initial recombination produced by the ionizing radiation per pulse. Considering the presence of free electrons in the chamber, the alternative analytical models for the CCE are:

\begin{equation}
\text{CCE}_{\text{II}} =  \frac{1}{u} \log\left[1+\frac{e^{pu}-1}{p}\right] 
\end{equation}

\begin{equation}
\text{CCE}_{\text{III}} = p+ \frac{1}{u} \log\left[1+(1-p)\;u\right] 
\end{equation}

\begin{equation}
\text{CCE}_{\text{IV}} = \lambda + \frac{1}{u}\log\left[1 + \frac{e^{\lambda(1-\lambda)u} - 1}{\lambda}\right] \quad\quad\quad \lambda = 1-\sqrt{1-p}
\end{equation}

Recently, Fenwick and Kumar\citep{fenwick_collection_2022} solved the problem using the exact distribution of free electrons in the chamber for a constant unperturbed electric field. In their model, the CCE can be written as:
\begin{equation}
\text{CCE}_{\text{FK}} = \frac{1}{u} \log\left(1 + R \, \exp(R) \left[E_1\left(R \, \exp(-\frac{d}{v_e\,\tau})\right) - E_1(R)\right]\right);   \quad\quad\quad R = \frac{u v_e\tau}{d}
\end{equation}

\noindent where $E_1$ represents the exponential integral function. These four models include the effect of free electrons, but share other limitations of the original Boag's model (no electric field perturbation and instantaneous release of the charge).

\noindent On the other hand, the logistic model\citep{petersson_high_2017, bourgouin_charge_2023} has been proposed as a empirical description at high DPP of the CCE:

\begin{equation}
\text{CCE}_{\text{Logistic}} = \frac{1}{\left[1+ (\frac{D\,d^2}{U})^{A} \right]^{B}};
\end{equation}

\noindent where $A$ and $B$ are fit parameters, $D$ is the dose per pulse in mGy, $U$ is the applied voltage in V and $d$ is the distance between electrodes in mm. For the Advanced Markus operated at 300~V and for a 1.8~$\upmu$s pulse duration, Petersson \textit{et al.}\citep{petersson_high_2017} reported values ranging from 2.2 to 2.9 for parameter $A$ and from 0.119 to 0.169 for parameter $B$.

\section{Results}
\subsection{Calibration of the ICT by means of a flash diamond}
\begin{figure}[!t]
   \begin{center}
   \includegraphics[width=10cm]{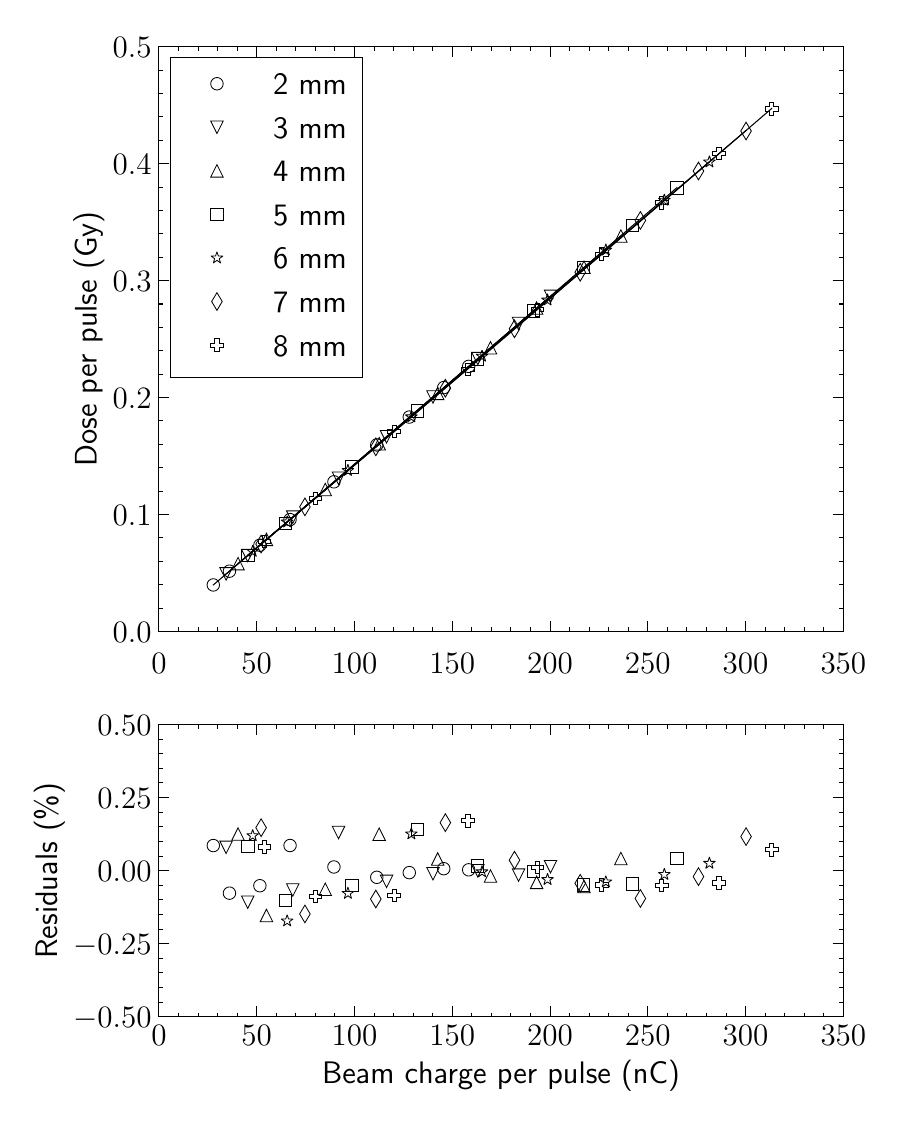}
   \captionv{15}{}{3rd order polynomial fit of the fD detector signal against the ICT charge per pulse for each slit opening when the pulse duration is changed.\label{fD7610_fit}}
    \end{center}
\end{figure}

\begin{figure}[!ht]
   \begin{center}
   \includegraphics[width=12cm]{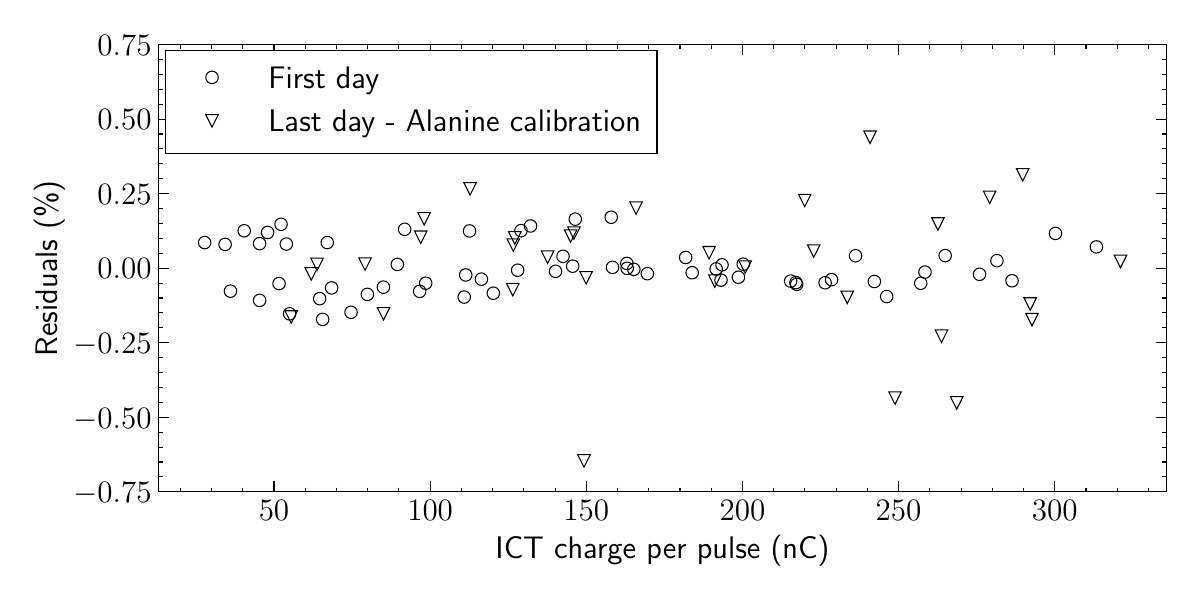}
   \captionv{15}{}{Residuals of the estimated fD charge per pulse using the 3rd order polynomial fits the first day and the last day, when the alanine calibration was performed.\label{fD7610_stab}}
    \end{center}
\end{figure}

The value for the calibration coefficient of the fD is 3.929~$\text{Gy\,nC}^{-1}$, obtained from the measurements against alanine. The reference depth of measurement for the electron beam quality used in this work was 4.10~g\,cm$^{-2}$, determined from depth dose curve measured with the fD. The 3rd order polynomial fit for each slit opening of the fD signal against the charge per pulse of the electron beam is shown in Figure~\ref{fD7610_fit}.
The presented residuals (Figure~\ref{fD7610_stab}) are relative to the polynomial fit done on the initial day of measurements. The standard deviation obtained for the last day is 0.22~\% which is taken into account in the uncertainties as a evaluation of the short-term stability of the linear accelerator.

\subsection{Uncertainty budget for the charge collection efficiency}
The uncertainty estimation of the collected charge from the IC was obtained as the standard deviation of the average of 25 pulses once corrected by the charge per beam pulse. The uncertainty due to deviation of the actual dose from the value determined via the ICT using the polinomial fit was estimated using the uncertainty propagation taking into account the covariance between the fitted parameters. The list of contributions considered for the estimation of the combined uncertainty for the CCE measurement is given in Table~\ref{unc_list}.

\begin{table}[!b]
\begin{center}
\captionv{15}{}{List of the sources of uncertainty and its average and range values considered for the calculation of the uncertainty of the CCE.\label{unc_list}\label{unc_list} \vspace*{2ex}}
\begin{tabular} {lccc}
\hline
\hline
\textbf{Uncertainty source} & \textbf{Type} & \textbf{Average} & \textbf{Range} \\
\hline
$Q$                        & A & 0.02~\% & 0.01~\% - 0.12~\%   \\
$k_{\text{elec}}$          & B & 0.10~\%  &                     \\
$k_{\text{Q}, \text{Q}_0}$ & B & 0.70~\%  &                     \\
$k_{\text{TP}}$            & B & 0.05~\% &                     \\
$N_{\text{D,w,Q}_0}$       & B &         & 0.25~\% - 0.40~\%    \\
$D_{\rm ref}$                  & B & 0.77~\% & 0.77~\% - 1.0~\%    \\
Positioning                & B & 0.07~\% &                     \\
Stability                  & A & 0.22~\% &                     \\
\hline
Combined                   &   & 1.1~\% & 1.1~\% - 1.3~\%   \\
\hline
\hline
\end{tabular}
\end{center}
\end{table}
\subsection{Comparison of the analytical methods\label{S_comp}}
\begin{figure}[!h]
   \begin{center}
   \includegraphics[width=9cm]{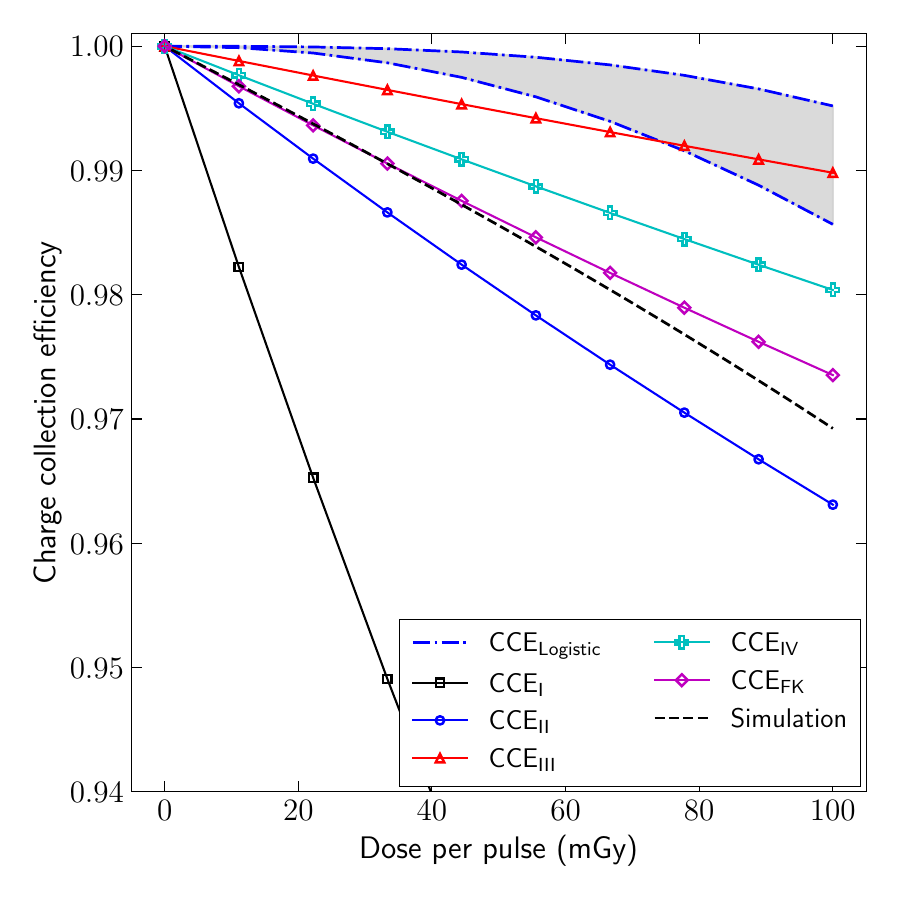}
   \captionv{15}{}{Comparison of the logistic model from Petersson \textit{et al.}\citep{petersson_high_2017} using the reported range of fit values (CCE$_{\text{Logistic}}$), the different Boag models (CCE$_{\text{I}}$ to CCE$_{\text{IV}}$) and the Fenwick and Kumar model\citep{fenwick_collection_2022} (CCE$_{\text{FK}}$) corresponding to the models outlied in Section~\ref{AModels} and the simulation presented in Section~\ref{NumModel} for a PPIC of 1~mm gap (as e.g. Advanced Markus) operated at 300~V and exposed to a 1.8~$\upmu$s pulse duration time.\label{AnalyticalComp}}
    \end{center}
\end{figure}

The models outlined in Section~\ref{AModels} are compared in Figure~\ref{AnalyticalComp} for a PPIC of 1~mm gap (as e.g. Advanced Markus) operated at 300~V and exposed to a 1.8~$\upmu$s pulse duration time. The transport parameters used in both the simulation and the analytical expressions correspond to those specified in Table~\ref{trasp_parameters}. Furthermore, the calibration coefficient of the Advanced Markus 2320 was used for the calculation of the released charge in the medium for a given DPP to water. For the logistic model, a region delimited by the values of the parameters reported by Petersson~\textit{et al.}\citep{petersson_high_2017} is presented in Figure~\ref{AnalyticalComp}. This exhibits the significant differences in the prediction of the different analytical models for a DPP interval between 0 and 100 mGy. It should be noted that the Petersson~\textit{et al.} fit was optimized using data up to 10 Gy.\\
\subsection{Charge collection efficiency and polarity correction factor\label{S_CCE}}
\begin{figure}[p]
   \begin{center}
   \includegraphics[width=15cm]{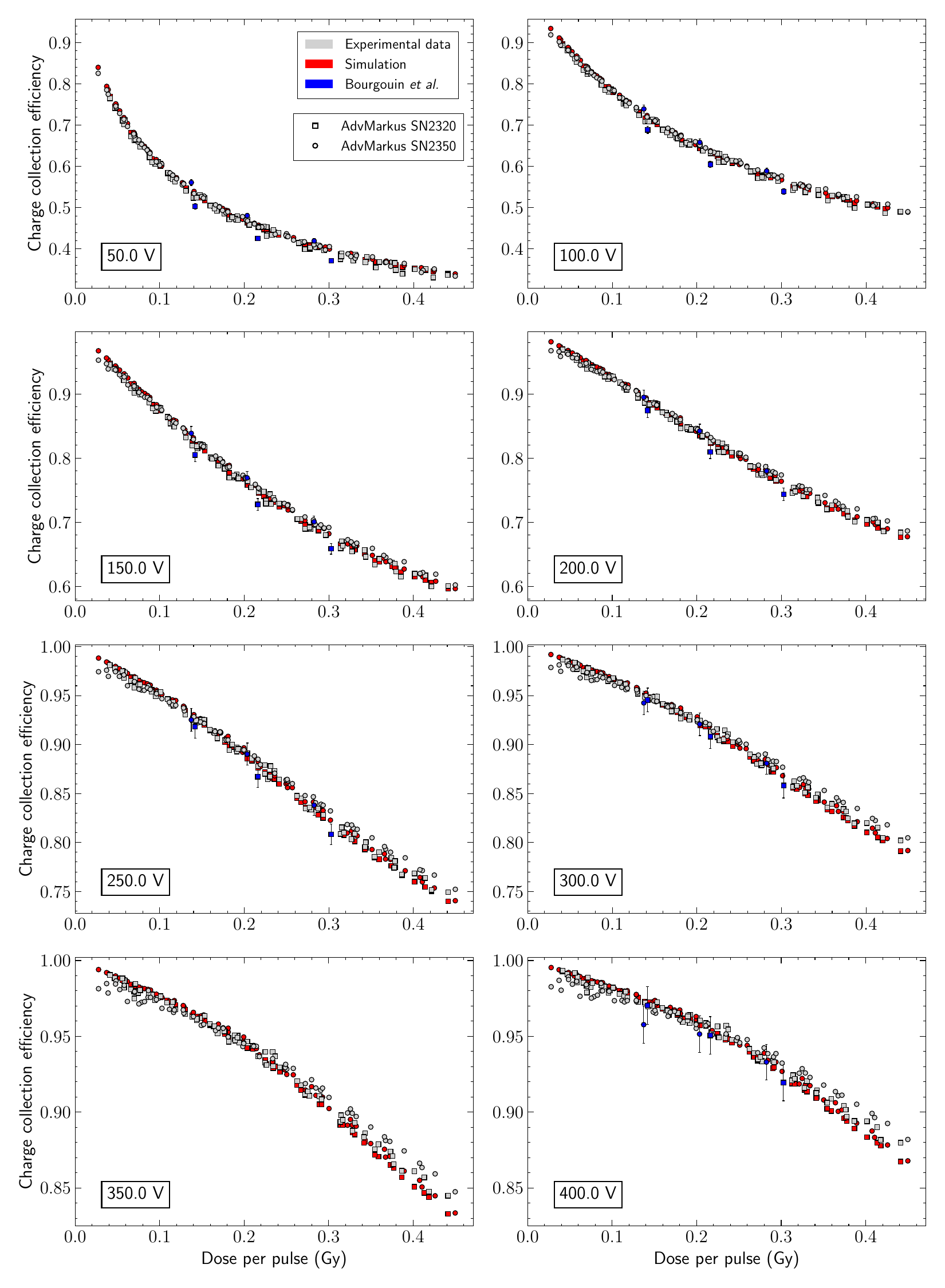}
   \captionv{15}{}{Experimental determined CCE for the two Advanced Markus chambers measured as a function of the DPP. Each panel contains the obtained CCE for a different operating voltage. \label{AdvM_CCE}}
    \end{center}
\end{figure}

\begin{figure}[p]
   \begin{center}
   \includegraphics[width=15cm]{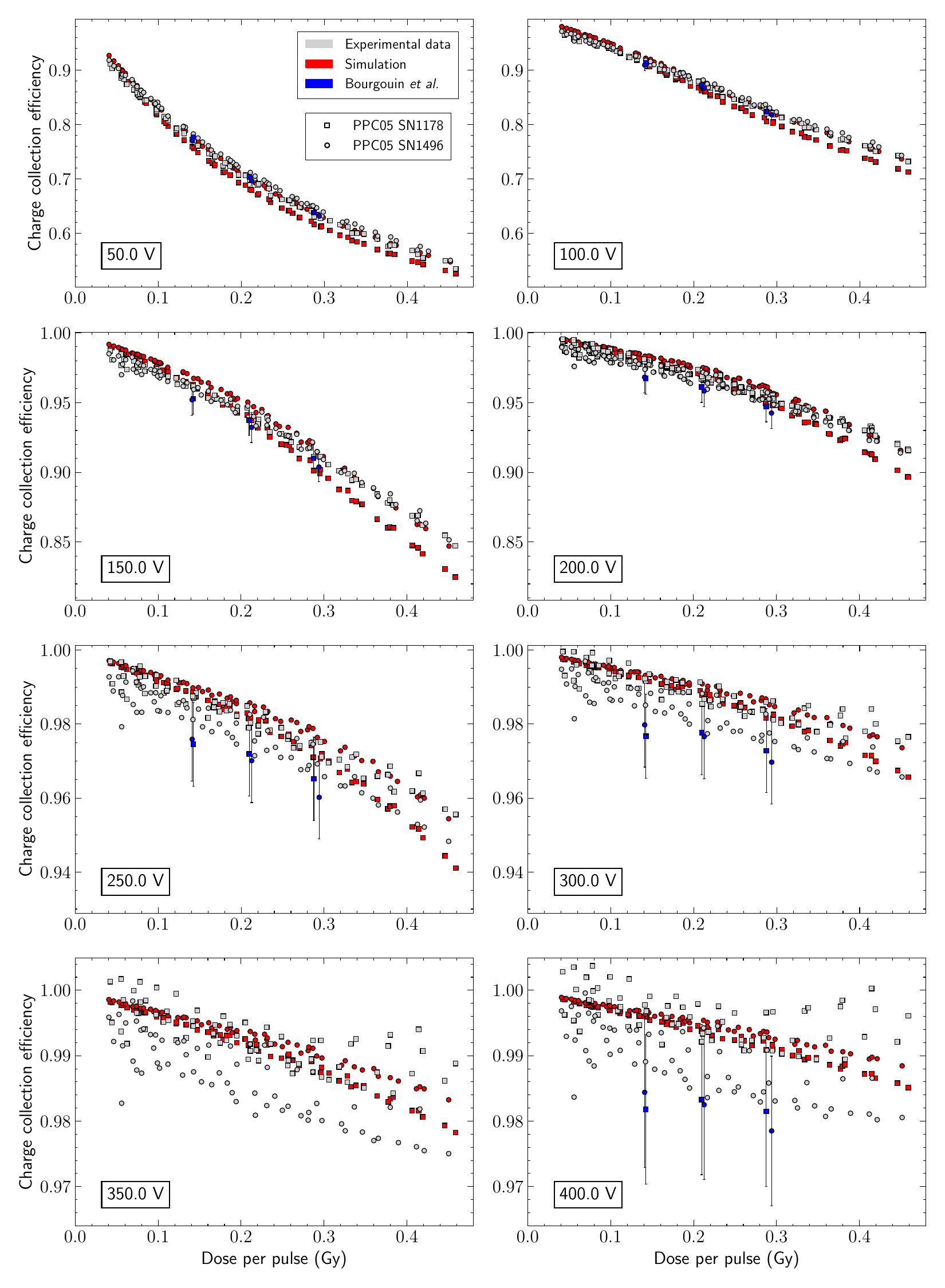}
   \captionv{15}{}{CCE for the two PPC05 ionization chambers investigated for different polarization voltages as a function of the DPP.\label{PPC05_CCE}}
    \end{center}
\end{figure}

\begin{figure}[!t]
   \begin{center}
   \includegraphics[width=17cm]{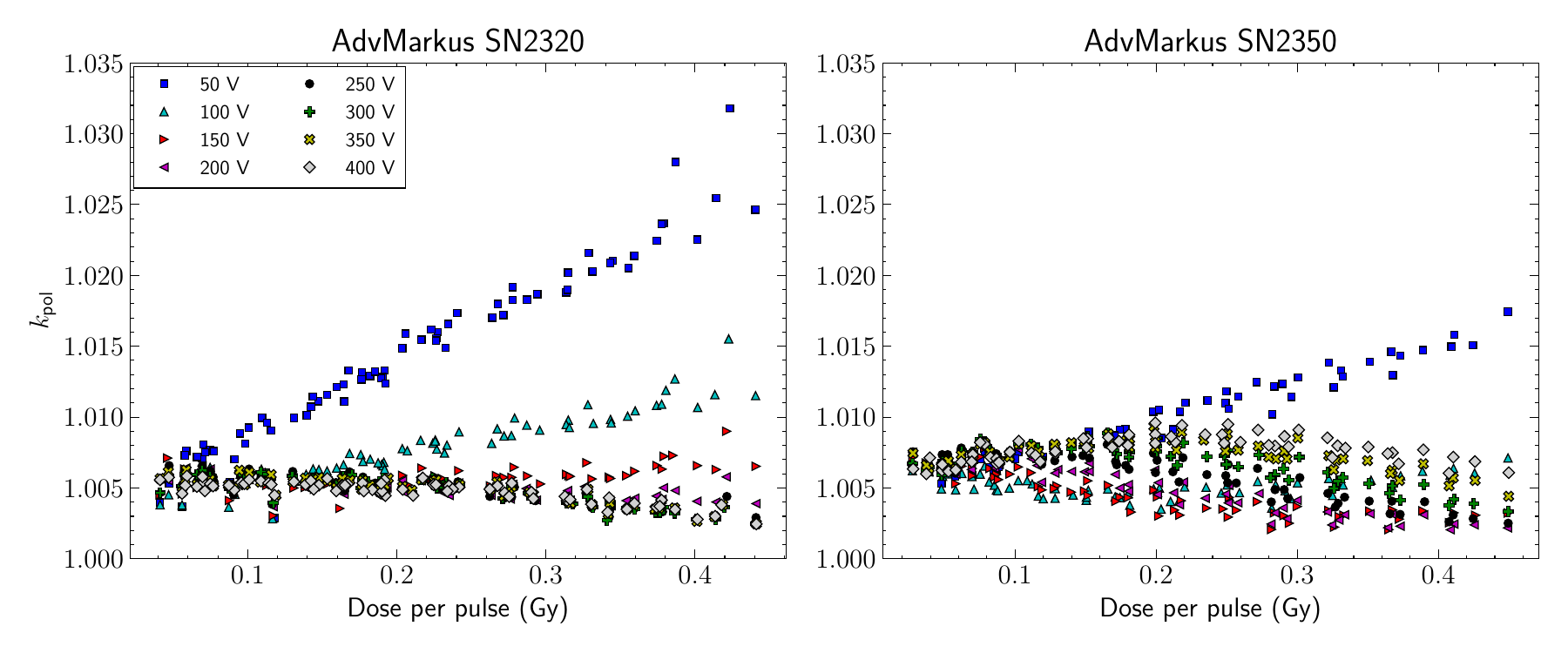}
   \captionv{15}{}{Polarity correction factor obtained according to equation~\ref{kpol_eq} for the two Advanced Markus chambers investigated as a function of the DPP. Each color in the graphic represents a different value of the polarization voltage\label{AdvM_polarity}}
    \end{center}
\end{figure}

\begin{figure}[!t]
   \begin{center}
   \includegraphics[width=17cm]{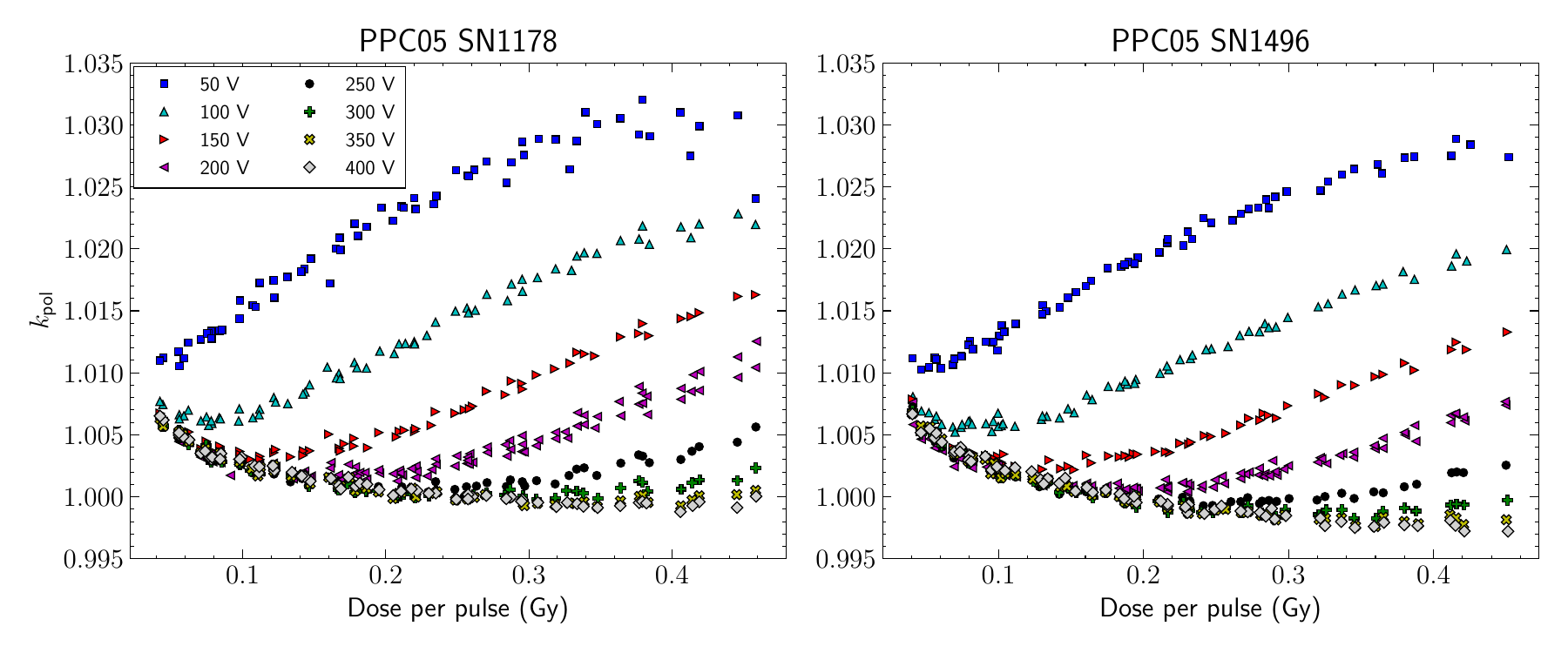}
   \captionv{15}{}{Polarity effect for the two PPC05 ionization chambers investigated for the different polarization voltages as a function of the DPP.\label{PPC05_polarity}}
    \end{center}
\end{figure}

The experimental determined CCE of the Advanced Markus PPICs and its comparison to the experiment of Bourgouin~\textit{et al.}\citep{bourgouin_charge_2023} and our numerical simulation are shown in Figure~\ref{AdvM_CCE} while the polarity effect is presented in Figure~\ref{AdvM_polarity}. Similarly, the obtained CCE for the PPC05 is presented in Figure~\ref{PPC05_CCE} and the polarity correction factor is shown in Figure~\ref{PPC05_polarity}. The uncertainty bars are only displayed for the data of Bourgouin~\textit{et al.} for the sake of clarity. The relative uncertainty of the CCE and its sources are listed in Table~\ref{unc_list}, while for the polarity correction factors, the estimated uncertainty is 0.23~\% on average.

\subsection{Two-voltage method\label{S_TVM}}
\begin{figure}[!ht]
   \begin{center}
   \includegraphics[width=16cm]{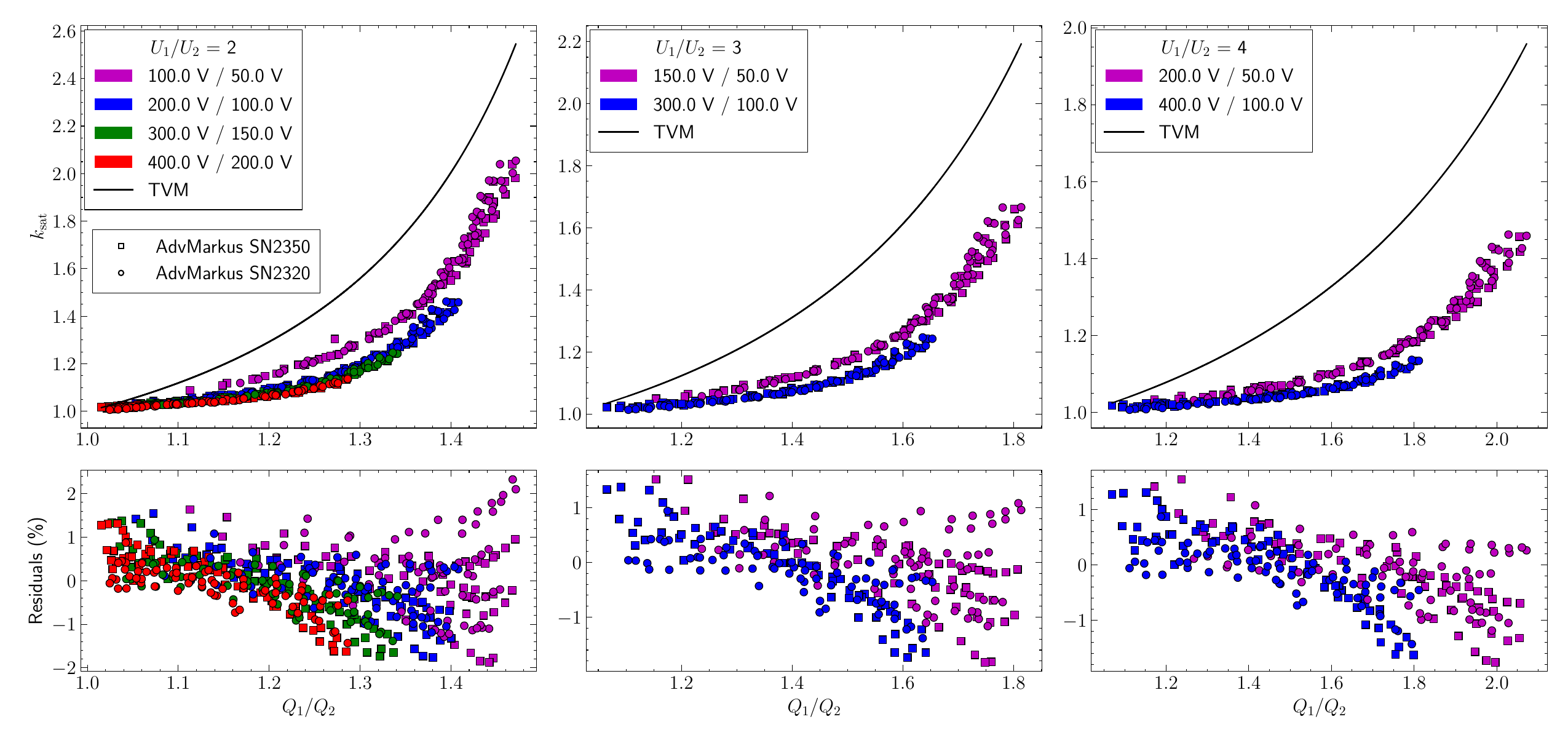}
   \captionv{15}{}{Saturation factor ($k_{\rm sat}$) for the two Advanced Markus PPIC investigated as a function of the experimental charge ratio. Each panel represents a different voltage ratio and each color is a different pair of voltages. The black solid line is the Boag TVM and the lower planes represent the residuals with respect to the saturation factor obtained with simulation. The dose per pulse was varied between 30~mGy up to 460~mGy.\label{TVM_AdvM}}
    \end{center}
\end{figure}

\begin{figure}[!ht]
   \begin{center}
   \includegraphics[width=16cm]{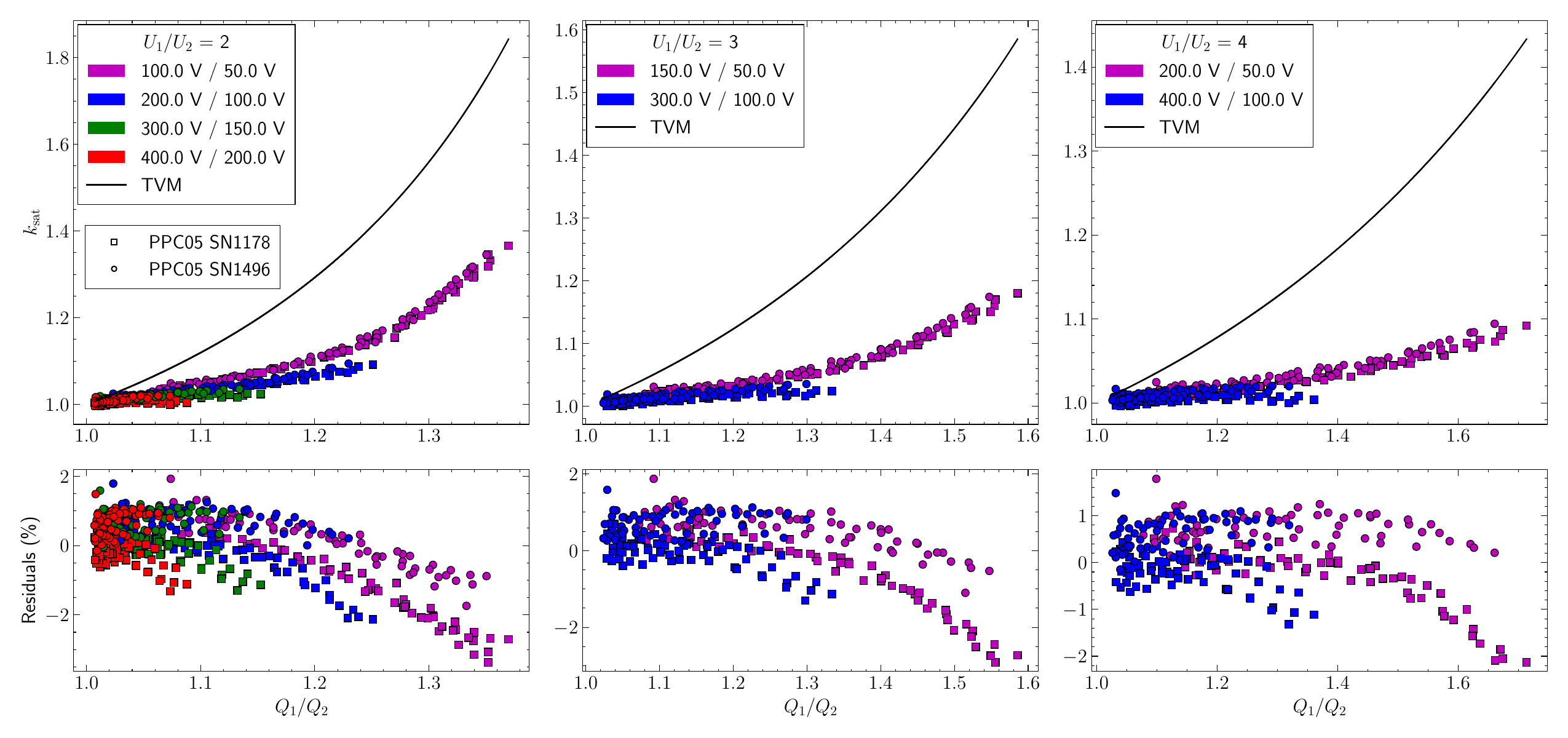}
   \captionv{15}{}{Saturation factor ($k_{\rm sat}$) for the two PPC05 PPIC investigated as a function of the experimental charge ratio. Each panel represents a different voltage ratio and each color is a different pair of voltages. The black solid line is the Boag TVM and the lower panels represent the residuals with respect to the saturation factor obtained with simulation.The dose per pulse was varied between 40~mGy up to 460~mGy.\label{TVM_PPC05}}
    \end{center}
\end{figure}

The saturation factor as a function of the charge ratios is presented in Figure~\ref{TVM_PPC05} for the PPC05 and in Figure~\ref{TVM_AdvM} for the Advanced Markus ionization chamber. In these figures, each panel represents a different voltage ratio that can be obtained with a different choice of voltage pairs that are represented in different colors. For comparison the continuous black line that represents the TVM based on Boag's model is plotted against the collected charge ratio. This line is obtained solving the following transcendental equation for a given collected charges $Q_1$ and $Q_2$ at voltages $U_1$ and $U_2$ ($U_1>U_2$):
\begin{equation*}
\frac{Q_1}{Q_2} = \frac{U_1}{U_2}\frac{\log\left(1 + u\right)}{\log\left(1 + \frac{U_1}{U_2}u\right)}
\end{equation*}
Additionally, in the lower part of the figures, the residuals of the experimental data with respect to the simulation results are shown. 

\subsection{Polynomial fit of the numerical solutions\label{S_fit}}
The CCE obtained with the numerical simulation was fitted as a function of the collected charge  using a 3rd order polynomial for a 1 mm and a 0.6 mm electrode distance PPIC. The obtained coefficients are tabulated as a function of the voltage ratio, $n=\frac{U_1}{U_2}$, and the pulse duration $\Delta t$:
\begin{equation}
k_{\rm sat} = a(\Delta t, n)\, x^3 + b(\Delta t, n) \, x^2 + c(\Delta t, n)\, x + d(\Delta t, n), \quad\quad\quad x = \frac{Q_{1}}{Q_{2}}
\label{pol_fit}
\end{equation}
where $Q_1$ is the collected charge at the nominal voltage of operation and $Q_2$ is the collected charge for the lower voltage. The coefficient $d$ satisfies the relation $d = 1 - (a + b + c)$ to ensure that the CCE is 1 when the charge ratio is 1. In all the simulations used to obtain this fit, temperature and pressure were fixed to standard conditions, 1013.25~hPa and 20~$^\circ$C, respectively. 

These simulations were conducted for pulse durations ranging from 0.5~$\upmu$s up to 5.0~$\upmu$s, which are typical values for conventional clinical and preclinical linear accelerators. For these fits, at least 100 equally spaced points in released charge per pulse were simulated yielding saturation factors from 1.002 to 1.055.

Although the TRS-398 recommends the use of voltage ratios $n=\frac{U_1}{U_2}\geq3$, we decided to perform the polynomial fits using voltages with a ratio $n=2$. This choice mitigates the reported effect of an increase in charge when increasing the voltage, as noted by some authors in previous publications\cite{deblois_saturation_2000, palmans_ion_2010,rossomme_correction_2021}. It is worth to mention that the inclusion of the free electron fraction reduces the sensitivity of the saturation factor parametrizations with respect to the charge ratio compared to the classical TVM method, where the voltage ratio equal to 2 is usually disregarded. 

\begin{figure}[!t]
   \begin{center}
   \includegraphics[width=10cm]{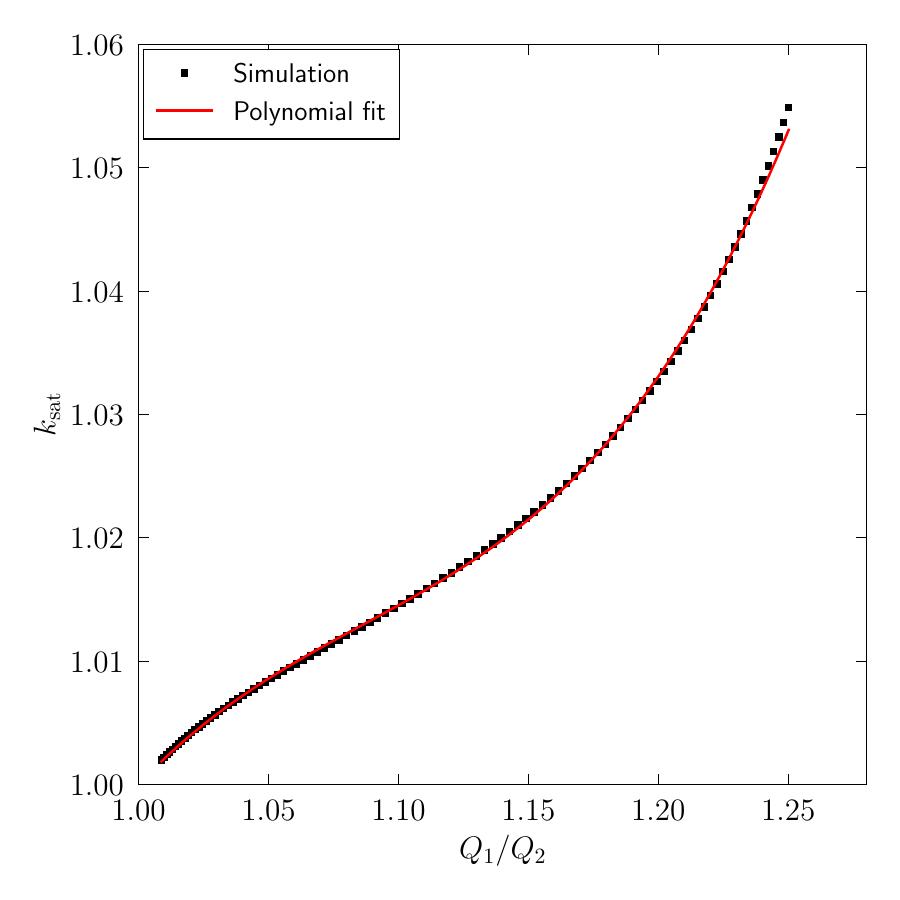}
   \captionv{15}{}{Example of the polynomial fit performed for a 0.6 mm PPIC and voltage ratio of 300~V/150~V. The points represent the 100 simulated points with a saturation factor of between 1.002 and 1.055 and the continous line the 3rd order polynomial fit.\label{PFitExample}}
    \end{center}
\end{figure}

Figure~\ref{PFitExample} shows an example of the polynomial fit corresponding to a 0.6 mm PPIC and a voltage quotient of 300~V/150~V. The obtained coefficients are presented in Table~\ref{PPC05_coefs} for the 0.6 mm PPIC and in Table~\ref{AdvM_coefs} for the 1.0~mm PPIC. Preliminary results show that the relative uncertainty component due to the charge transport coefficients and deviation from the polynomial fit can amount up to a 0.5~\%. However, further investigation is required to evaluate the existing correlation between the different transport parameters and volume recombination.

\begin{figure}[!t]
   \begin{center}
   \includegraphics[width=14cm]{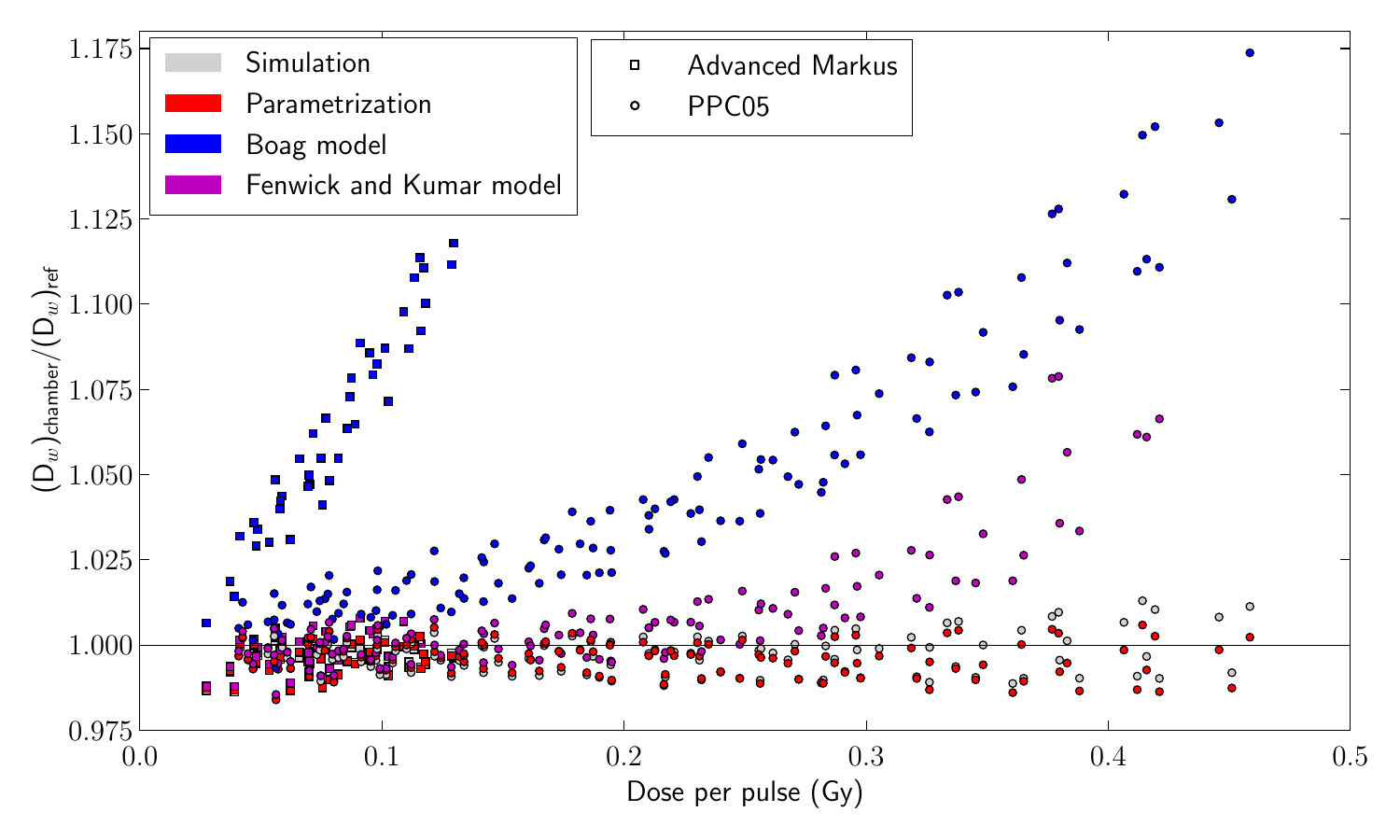}
   \captionv{15}{}{Ratio of the dose per pulse obtained with the ionization chamber corrected using Boag model, Fenwick and Kumar model, simulation and the parametrization from Tables~\ref{PPC05_coefs}~and~\ref{AdvM_coefs} over the reference dose per pulse. The pair of voltages chosen here is 300~V and 150~V and only the data with charge ratios where the parametrization gives $k_{\rm sat}<$~1.05 is used for this comparison. For the model of Fenwick and Kumar only charge ratios lower than 1.12 were considered as the convergence of the solution to the transcendental equation was poor because of the high sensibility to the charge ratio. \label{POC}}
    \end{center}
\end{figure}

As a proof of consistency, Figure~\ref{POC} represents the quotient of the dose per pulse obtained with the ionization chamber over the reference dose per pulse corrected by ion recombination using the Boag formula, the Fenwick and Kumar formula, the numerical simulation and the presented parametrization of the data that satisfy the condition $k_{\rm sat}<$~1.05. It is worth noting that the experimental ratio of collected charges was corrected by the dose per pulse due to accelerator fluctuations during the beam delivery. The limit of validity of these polynomials is achieved when the obtained $k_{\rm sat}$ is larger than 1.05.

\begin{table}[!b]
\begin{center}
\vspace*{2ex}
\captionv{15}{}{Fitting parameters for the calculation of the saturation factor for a 0.6 mm PPIC according to equation \ref{pol_fit} valid in the range $k_{\rm sat}<$~1.05. \label{PPC05_coefs} \vspace*{2ex}}
\begin{tabular} {ccccccccccc}
\hline\hline
 \multicolumn{2}{c}{ 0.6~mm } & \multicolumn{4}{c}{ 300~V / 150~V } & & \multicolumn{4}{c}{ 500~V / 250~V } \\
\cline{1-1} \cline{3-6} \cline{8-11}
\makecell{Pulse \\ duration ($\upmu$s)} & & a & b & c & d & & a & b & c & d \\\hline
 0.50 & & 9.685 & -31.159 & 33.584 & -11.110 & & 13.333 & -43.088 & 46.496 & -15.741 \\
 1.00 & & 8.780 & -28.293 & 30.556 & -10.043 & & 10.682 & -34.648 & 37.540 & -12.574 \\
 2.00 & & 7.387 & -23.880 & 25.882 & -8.389 & & 7.967 & -25.985 & 28.318 & -9.300 \\
 3.00 & & 6.310 & -20.458 & 22.245 & -7.097 & & 6.224 & -20.394 & 22.336 & -7.166 \\
 4.00 & & 5.491 & -17.850 & 19.466 & -6.107 & & 5.154 & -16.954 & 18.644 & -5.844 \\
 5.00 & & 4.869 & -15.866 & 17.348 & -5.351 & & 4.473 & -14.769 & 16.297 & -5.001 \\
\hline\hline
\end{tabular}
\end{center}
\end{table}
\begin{table}[!t]
\vspace*{2ex}
\begin{center}
\captionv{15}{}{Fitting parameters for the calculation of the saturation factor for a 1.0 mm PPIC according to equation \ref{pol_fit} valid in the range $k_{\rm sat}<$~1.05.\label{AdvM_coefs}\vspace*{2ex}}
\begin{tabular} {ccccccccccc}
\hline\hline
 \multicolumn{2}{c}{ 1.0~mm } & \multicolumn{4}{c}{ 300~V / 150~V } & & \multicolumn{4}{c}{ 400~V / 200~V } \\
\cline{1-1} \cline{3-6} \cline{8-11}
\makecell{Pulse \\ duration ($\upmu$s)} & & a & b & c & d & & a & b & c & d \\\hline
 0.50 & & 3.263 & -10.110 & 10.766 & -2.919 & & 4.521 & -14.329 & 15.435 & -4.627 \\
 1.00 & & 3.214 & -9.969 & 10.625 & -2.870 & & 4.447 & -14.109 & 15.213 & -4.551 \\
 2.00 & & 3.145 & -9.770 & 10.427 & -2.802 & & 4.291 & -13.636 & 14.724 & -4.379 \\
 3.00 & & 3.078 & -9.578 & 10.236 & -2.736 & & 4.133 & -13.153 & 14.221 & -4.201 \\
 4.00 & & 3.012 & -9.388 & 10.045 & -2.669 & & 3.978 & -12.679 & 13.727 & -4.026 \\
 5.00 & & 2.963 & -9.249 & 9.909 & -2.623 & & 3.826 & -12.214 & 13.241 & -3.853 \\
\hline\hline
\end{tabular}
\end{center}
\end{table}

\section{Discussion}
The different existing analytical models for recombination yield different values of CCE even for moderate ($<$~30~mGy) dose per pulse as shown in Figure~\ref{AnalyticalComp}. Among these models, the first Boag model, CCE$_{\text{I}}$, consistently yields the lowest CCE values compared to all the other models that include the free electron contribution. This observation aligns with the fact that excluding free electrons in the model results in a higher negative ion density and an increased ion-ion recombination. As a consequence, the primordial Boag model likely tends to overestimate the saturation factor\cite{laitano_charge_2006} even for PPIC with a distance between electrodes relatively large (e.g. Roos or PPC40 with 2~mm gap) where the free electron fraction may certainly reach 20~\% at the recommended operational bias voltage\citep{paz-martin_numerical_2022}.

Boag's models that consider the free electrons' effect do not converge with the same slope when the DPP becomes lower (see Figure~\ref{AnalyticalComp}). The cause of this behaviour is due to the different way of including the free electron distribution between the different Boag models across the ionization chamber. This impacts the slope of the models when the DPP tends to zero. On the other hand, the Fenwick and Kumar model seems to agree with the simulation when the DPP is low enough: an agreement better than 0.3~\% is found when the DPP is lower than 85~mGy per pulse for the parameters of Figure~\ref{AnalyticalComp}. Moreover this is also confirmed when compared with the experimental data in Figure~\ref{POC}. Above this DPP of 85~mGy, differences are primarily attributable to electric field perturbation due to the high amount of charge density released by the pulse inside the chamber volume. Regarding the logistic model proposed by Petersson \textit{et al.}, it is not suitable for the DPP used in this work since the slope of the model does not reproduce the experimental data. CCE vs DPP slope tends to be null when approaching the very low DPP. This is expected since the model was designed for measurements at much higher dose per pulse than the used in this work.

\begin{figure}[!t]
   \begin{center}
   \includegraphics[width=9cm]{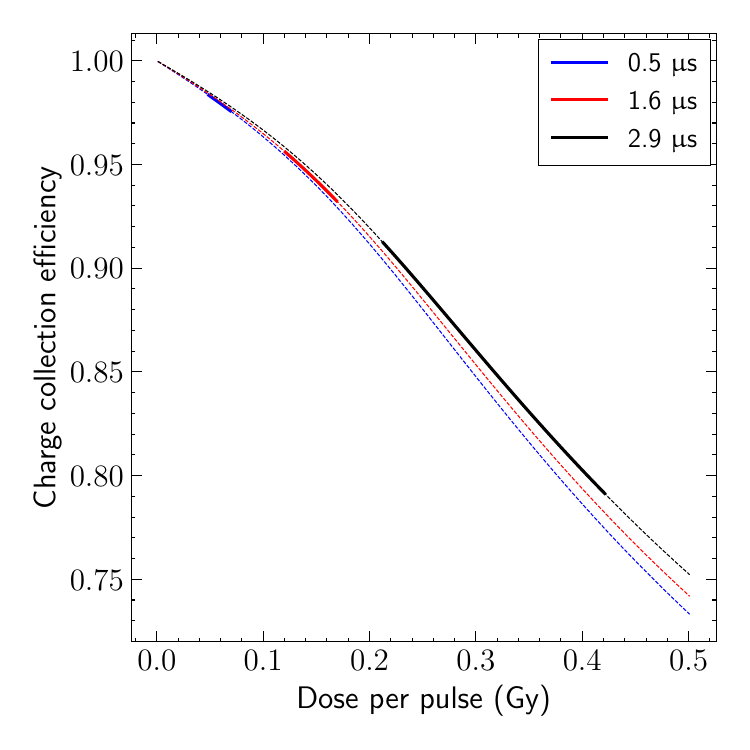}
   \captionv{12}{}{Comparison of the simulated CCE of an Advanced Markus-like PPIC at 300 V for different pulse durations in the same range of DPP studied in this work. While the dotted line represents the CCE over the entire range, the solid line represents the actual range of DPP covered for each pulse duration.\label{AdvM_pulsed}}
    \end{center}
\end{figure}

In this work, no significant dependency of the CCE with the pulse duration has been observed in the DPP range studied. On the other hand, whenever the same DPP is delivered with different pulse duration, the expected effect on the CCE would become more pronounced when the pulse duration is close to the charge collection time\citep{paz-martin_numerical_2022} (e.g. for a 1~mm PPIC operated at 400~V the charge collection time is around 20~$\upmu$s). The specific way the DPP is varied in the set-up considered in this work limits significantly the sensitivity of the results to the pulse duration delivery. In this work, the instantaneous dose rate accessible is limited, producing a correlation between the pulse duration and DPP in the experimental data. This yields a non-uniform distribution of the experimental points, accumulating low pulse duration points at lower DPP and higher pulse duration points at higher DPP values. A graphical example of this behavior, obtained using simulation for our specific experimental measurement setup, is shown in Figure~\ref{AdvM_pulsed}, where the solid line represents the actual range of DPP covered in this work, and the dotted line shows how it varies outside this range for different pulse duration.

In other words, there is not enough overlap in the DPP delivered between the different pulse durations used and we cannot make a conclusive statement about the pulse duration just attending to the experimental data presented. Nevertheless, the experimental results are compatible within uncertainties with the predicted results of numerical simulation taking into account the pulse duration. Consequently, we have included the pulse duration in the parametrization of the saturation factors in Tables \ref{PPC05_coefs} and \ref{AdvM_coefs}. 

\begin{figure}[!t]
\centering
\includegraphics[width = 9 cm]{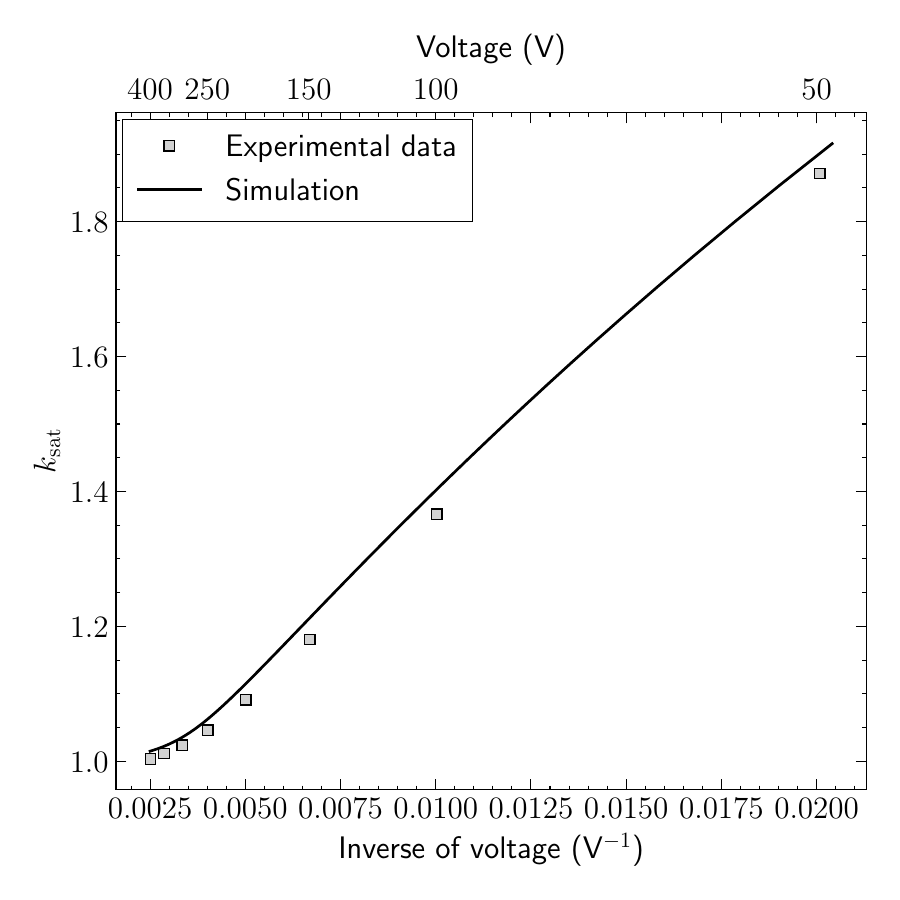}
\captionv{12}{}{Relation of the inverse of the charge (here presented as saturation factor) versus the inverse of the voltage for one of the PPC05 ionization chamber irradiated with 0.45~Gy per pulse and a 2.9~$\upmu$s pulse duration.\label{diagLin}}
\end{figure}

Although the original Boag's model predicts that the charge quotient of a chamber would only depend on the voltage ratio, Figure~\ref{TVM_PPC05} and Figure~\ref{TVM_AdvM} show clearly that different pairs of voltages with equal voltage ratio exhibit different relationship between charge quotient $\frac{Q_1}{Q_2}$ and saturation factor $k_{\rm sat}$. The TVM method based on the Boag model also predicts that the distance between electrodes should not have any significant impact on the saturation correction factor evaluation. However, the observed dependence of the charge ratio for the saturation factor of Advanced Markus and PPC05 chambers exhibit significant differences that are also translated to the parametrization included in this work. Also, the linear extrapolation using the Jaffe diagram is compromised at higher voltages as seen in Figure~\ref{diagLin} and previously reported by Fenwick and Kumar\citep{fenwick_collection_2022}.

Additionally, it has been shown by other authors that there is an increase in the chamber collected charge with the applied voltage not related to volume recombination effect\cite{deblois_saturation_2000,palmans_ion_2010,rossomme_correction_2021}. This will causes a more prominent effect in the TVM whenever the ratio of two voltages used becomes larger. In order to minimize this contribution, the ratio of voltages selected in this work in the saturation correction factor parametrization has been kept to a value of 2.

In a recent publication of Bourgouin \textit{et al.}\citep{bourgouin_charge_2023}, a discrepancy between measurements performed between the Federal Office of Metrology of Switzerland (METAS) and PTB of CCE was observed. These discrepancies were attributed to the large difference (around 40~hPa) in the absolute pressure of the air during the measurements in combination with other factors. During the measurements performed in this work, the average temperature and pressure were 18.55(20)~$^\circ$C and 1003.7(5.4)~hPa, where the number in brackets is the standard deviation with k = 1. The possible variation of CCE due to pressure and temperature of data for the average values is estimated by simulation to be less than 0.2~\% on average and 0.9~\% on maximum, with respect to the average values.

Considering the results that show the explicit dependence of the charge ratio with chosen voltages (with a fixed ratio of 2), we have performed a polynomial parametrization of $k_{\rm sat}$ up to 3rd order for 500~V/250~V and 300~V/150~V for the PPC05 chamber and 300~V/150~V and 400~V/200~V for the Advanced Markus chamber.

\section{Conclusions}

The CCE of two commercial ionization chambers, namely PPC05 and Advanced Markus, was measured varying the DPP between 30~mGy per pulse and 460~mGy per pulse and pulse duration ranging between 0.5~$\upmu$s and 2.9~$\upmu$s. The experimental data was obtained using the PTB electron research linear accelerator\citep{schuller_metrological_2019}. The polarity correction factor was also obtained and shown to depend on the DPP, in line with recent investigations.

Variations of the CCE for a given value of DPP with the pulse duration have not been observed within the statistical uncertainties reported\citep{bourgouin_charge_2023}. This fact is explained in terms of the correlation between the pulse duration and the DPP achieved and is supported by the simulation. The charge ratios are found to be dependent on the actual pair of voltages $U_{1}$ and $U_{2}$ chosen even when the ratio is constant.

We have used a numerical method to evaluate the CCE from first principles\citep{paz-martin_numerical_2022}. Overall, a satisfactory agreement between the experimental data and the simulation data is found. The average discrepancy in terms of CCE for the PPC05 is 0.7~\% and for the Advanced Markus is 0.5~\%, while the maximum discrepancy is 4.6~\% and 3.4~\%, respectively.

Our experimental data suggest that the existing recommendation in the codes of practice to account for volume recombination systematically overestimates the saturation factor, especially when it becomes larger ($>$ 1.02). Moreover, variations with the used pair voltages and PPIC distance between electrodes are observed, which is in direct contradiction with the original TVM outcomes. 

As an alternative, 3rd order polynomial fits are given for the two PPICs studied in this work for voltage ratios of 2 to avoid effects such as the charge increase observed when voltage increases in previous publications, which is not yet completely understood. This parametrization of the saturation
factor is able to correct the recombination effect with an average deviation below 0.7~\% for the Advanced Markus and the PPC05 using 300~V/150~V and below 0.6~\% for the Advanced Markus using a voltage ratio of 400~V/200~V. It is worth to mention that for the PPC05 ionization chamber and the voltage ratio of 300~V/150~V a maximum k$_{\text{sat}}$ of 1.036 is reached for the range of dose per pulse investigated.

In summary, the classical TVM based on the Boag model that does not consider free electrons in the ionization chamber consistently overestimates the saturation correction factor. These results may have important implications for the dosimetry with ionization chambers in therapies that use a dose per pulse higher than conventional, such as intraoperative radiotherapy or FLASH modalities, but may also be important for the dosimetry of conventional techniques that employ lower DPP values.

\section*{Acknowledgments}
Jose Paz-Martín has received a predoctoral research contract from the Xunta de Galicia regional government. Juan Pardo-Montero acknowledges the support of Xunta de Galicia, Axencia Galega de Innovación (grant IN607D2022). This work has been developed under the project 18HLT04 UHDpulse with funding from the EMPIR innovation program. This work has received funding from `la Caixa' Foundation under the grant agreement HR23-00718 (`Dosimetry monitor for FLASH therapy' project) and from Grant PLEC2022-009476 funded by MCIN/AEI/10.13039/501100011033 and by the `European Union NextGenerationEU/PRTR'. The authors thank Christoph Makowski for the electron linear accelerator maintenance during the measurement campaign, Olaf Tappe for the machining work, and Thomas Hackel for the alanine measurements.

\section*{References}
\addcontentsline{toc}{section}{\numberline{}References}
\vspace*{-20mm}






\begin{thebibliography}{10}

\bibitem{agency_absorbed_2024}
International Atomic Energy Agency, Absorbed Dose Determination in External Beam Radiotherapy, Technical Reports Series No. 398 (Rev. 1), IAEA, Vienna (2024), doi:10.61092/iaea.ve7q-y94k

\bibitem{almond_aapms_1999}
Almond PR, Biggs PJ, Coursey BM, et al. AAPM's TG-51 protocol for clinical reference dosimetry of high-energy photon and electron beams. \textit{Med Phys.} 1999;26(9):1847-1870. doi:10.1118/1.598691

\bibitem{boag_ionization_1950}
Boag JW. Ionization measurements at very high intensities. Pulsed radiation beams. \textit{Br J Radiol.} 1950;23(274):601-611. doi:10.1259/0007-1285-23-274-601

\bibitem{boag_effect_1996}
Boag JW, Hochhäuser E, Balk OA. The effect of free-electron collection on the recombination correction to ionization measurements of pulsed radiation. \textit{Phys Med Biol.} 1996;41(5):885-897. doi:10.1088/0031-9155/41/5/005

\bibitem{fenwick_collection_2022}
Fenwick JD, Kumar S. Collection efficiencies of ionization chambers in pulsed radiation beams: an exact solution of an ion recombination model including free electron effects. \textit{Phys Med Biol}. 2022;68(1):10.1088/1361-6560/aca74e. doi:10.1088/1361-6560/aca74e

\bibitem{gotz_new_2017}
Gotz M, Karsch L, Pawelke J. A new model for volume recombination in plane-parallel chambers in pulsed fields of high dose-per-pulse. \textit{Phys Med Biol.} 2017;62(22):8634-8654. doi:10.1088/1361-6560/aa8985

\bibitem{kranzer_ion_2021}
Kranzer R, Poppinga D, Weidner J, et al. Ion collection efficiency of ionization chambers in ultra-high dose-per-pulse electron beams. \textit{Med Phys.} 2021;48(2):819-830. doi:10.1002/mp.14620

\bibitem{gomez_development_2022}
Gómez F, González-Castaño DM, Fernández NG, et al. Development of an ultra-thin parallel plate ionization chamber for dosimetry in FLASH radiotherapy. \textit{Med Phys.} 2022;49(7):4705-4714. doi:10.1002/mp.15668

\bibitem{paz-martin_numerical_2022}
Paz-Martín J, Schüller A, Bourgouin A, et al. Numerical modeling of air-vented parallel plate ionization chambers for ultra-high dose rate applications. \textit{Phys Med.} 2022;103:147-156. doi:10.1016/j.ejmp.2022.10.006

\bibitem{kranzer_charge_2022}
Kranzer R, Schüller A, Gómez Rodríguez F, et al. Charge collection efficiency, underlying recombination mechanisms, and the role of electrode distance of vented ionization chambers under ultra-high dose-per-pulse conditions. \textit{Phys Med.} 2022;104:10-17. doi:10.1016/j.ejmp.2022.10.021

\bibitem{petersson_high_2017}
Petersson K, Jaccard M, Germond JF, et al. High dose-per-pulse electron beam dosimetry - A model to correct for the ion recombination in the Advanced Markus ionization chamber. \textit{Med Phys.} 2017;44(3):1157-1167. doi:10.1002/mp.12111

\bibitem{bourgouin_charge_2023}
Bourgouin A, Paz-Martín J, Gedik YC, et al. Charge collection efficiency of commercially available parallel-plate ionisation chambers in ultra-high dose-per-pulse electron beams. \textit{Phys Med Biol.} 2023;68(23):10.1088/1361-6560/ad0a58. doi:10.1088/1361-6560/ad0a58

\bibitem{bourgouin_characterization_2022}
Bourgouin A, Knyziak A, Marinelli M, Kranzer R, Schüller A, Kapsch RP. Characterization of the PTB ultra-high pulse dose rate reference electron beam. \textit{Phys Med Biol.} 2022;67(8):10.1088/1361-6560/ac5de8. doi:10.1088/1361-6560/ac5de8

\bibitem{schuller_metrological_2019}
Schüller A, Pojtinger A, Meier M, Makowski C, and Kapsch RP,
\newblock The Metrological Electron Accelerator Facility ({MELAF}) for Research in Dosimetry for Radiotherapy. {\em World Congress on Medical Physics and Biomedical Engineering 2018}, {IFMBE} Proceedings, pages 589--593, Springer.

\bibitem{marinelli_design_2022}
Marinelli M, Felici G, Galante F, et al. Design, realization, and characterization of a novel diamond detector prototype for FLASH radiotherapy dosimetry. \textit{Med Phys.} 2022;49(3):1902-1910. doi:10.1002/mp.15473

\bibitem{kranzer_response_2022}
Kranzer R, Schüller A, Bourgouin A, et al. Response of diamond detectors in ultra-high dose-per-pulse electron beams for dosimetry at FLASH radiotherapy [published correction appears in Phys Med Biol. 2022 Jul 08;67(14):]. \textit{Phys Med Biol.} 2022;67(7):10.1088/1361-6560/ac594e. doi:10.1088/1361-6560/ac594e

\bibitem{bourgouin_absorbed-dose--water_2022}
Bourgouin A, Hackel T, Marinelli M, Kranzer R, Schüller A, Kapsch RP. Absorbed-dose-to-water measurement using alanine in ultra-high-pulse-dose-rate electron beams. \textit{Phys Med Biol.} 2022;67(20):10.1088/1361-6560/ac950b. doi:10.1088/1361-6560/ac950b

\bibitem{mvoros_relative_2012}
Vörös S, Anton M, Boillat B. Relative response of alanine dosemeters for high-energy electrons determined using a Fricke primary standard. \textit{Phys Med Biol.} 2012;57(5):1413-1432. doi:10.1088/0031-9155/57/5/1413

\bibitem{muir_monte_2014}
Muir BR, Rogers DW. Monte Carlo calculations of electron beam quality conversion factors for several ion chamber types. \textit{Med Phys.} 2014;41(11):111701. doi:10.1118/1.4893915

\bibitem{biagi_monte_1999}
Biagi SF. Monte Carlo simulation of electron drift and diffusion in counting gases under the influence of electric and magnetic fields \textit{Nucl. Instrum. Methods Phys. Res. A}. 1999;421(1-2):234-240. doi:10.1016/S0168-9002(98)01233-9

\bibitem{zhang_prediction_2019}
Zhang B; He J; Ji Y. Prediction of average mobility of ions from corona discharge in air with respect to pressure, humidity and temperature. \textit{IEEE Trans. Dielectr. Electr. Insul.} 2019;26(5):1403-1410. doi: 10.1109/TDEI.2019.008001.

\bibitem{boissonnat_measurement_2016}
Boissonnat~G., Fontbonne~J.-M., J.~Colin, A.~Remadi, and S.~Salvador,
\newblock Measurement of ion and electron drift velocity and electronic attachment in air for ionization chambers. 2016. doi: 10.48550/arXiv.1609.03740

\bibitem{deblois_saturation_2000}
DeBlois F, Zankowski C, Podgorsak EB. Saturation current and collection efficiency for ionization chambers in pulsed beams. \textit{Med Phys.} 2000;27(5):1146-1155. doi:10.1118/1.598992

\bibitem{palmans_ion_2010}
Palmans H, Thomas RA, Duane S, Sterpin E, Vynckier S. Ion recombination for ionization chamber dosimetry in a helical tomotherapy unit. \textit{Med Phys.} 2010;37(6):2876-2889. doi:10.1118/1.3427411

\bibitem{rossomme_correction_2021}
Rossomme S, Lorentini S, Vynckier S, et al. Correction of the measured current of a small-gap plane-parallel ionization chamber in proton beams in the presence of charge multiplication. \textit{Z Med Phys.} 2021;31(2):192-202. doi:10.1016/j.zemedi.2021.01.008

\bibitem{laitano_charge_2006}
Laitano RF, Guerra AS, Pimpinella M, Caporali C, Petrucci A. Charge collection efficiency in ionization chambers exposed to electron beams with high dose per pulse. \textit{Phys Med Biol.} 2006;51(24):6419-6436. doi:10.1088/0031-9155/51/24/009

\bibitem{bourgouin_the_2023}
Bourgouin A, Hackel T, Kapsch RP. The PTB water calorimeter for determining the absolute absorbed dose to water in ultra-high pulse dose rate electron beams. \textit{Phys Med Biol.} 2023;68(11). doi:10.1088/0031-9155/51/24/009

\bibitem{Bancheri_A_2024}
Bancheri J, Seuntjens J. A semi-analytical procedure to determine the ion recombination correction factor in high dose-per-pulse beams. \textit{Med Phys.} 2024;51(6):4458-4471. doi:10.1002/mp.17005

\end{thebibliography}



\bibliographystyle{./medphy.bst}    


\end{document}